# Bound States in Nanoscale Graphene Quantum Dots in a Continuous Graphene Sheet

Jia-Bin Qiao[1], Hua Jiang[2], Haiwen Liu[1], Hong Yang[1], Ning Yang[1], Kai-Yao Qiao[1], and Lin He[1,*]

[1] Center for Advanced Quantum Studies, Department of Physics, Beijing Normal University, Beijing, 100875, People's Republic of China
[2] College of Physics, Optoelectronics and Energy, Soochow University, Suzhou, 215006, People's Republic of China
*Correspondence and requests for materials should be addressed to L.H. (e-mail: helin@bnu.edu.cn).

**Considerable efforts have been made to trap massless Dirac fermions in graphene monolayer, but only quasi-bound states are realized in continuous graphene sheets up to now. Here, we demonstrate the realization of bound states in nanoscale graphene quantum dots (GQDs) in a continuous graphene sheet. The GQDs are electronically isolated from the surrounding continuous graphene sheet by circular boundaries, which are generated by strong coupling between graphene and substrate. By using scanning tunneling microscopy (STM), we observe single-electron charging states of the GQDs, seen as Coulomb oscillations in the tunneling conductance. Evolution of single-electron tunneling of the GQDs between the Coulomb blockade regime and the Coulomb staircase regime is observed by tuning the STM tip-sample distances. Spatial maps of the local electronic densities reveal concentric rings inside the GQDs with each ring corresponding to a single Coulomb oscillation of the tunneling spectra. These results indicate explicitly that the electrons are completely trapped inside the nanoscale GQDs.**

The confinement of electrons in graphene monolayer faces challenges due to Klein tunneling of massless Dirac fermions [1-3]. Although different recipes for confining the Dirac fermions in graphene have been suggested, only quasi-bound states with a finite trapping time are realized very recently in a continuous graphene sheet with well-defined circular p-n junctions [4-6], even with some extremely external conditions [7-9], such as high magnetic fields, or supercritical charges. Here, we show that bound states can be formed in graphene quantum dots (GQDs), which are part of a continuous graphene sheet, with circular boundary formed due to strong coupling with substrate. In our scanning tunneling microscopy (STM) measurements, we used single-electron charging effects to identify the GQDs in the continuous graphene sheet. The single-electron charging phenomena are quite ubiquitous in atoms, molecules, and small isolated conducting islands. They are treated as a clear signature that the studied object is isolated from the surrounding by tunnel barriers and the number of electrons residing on the object is quantized [10-16]. To observe the single-electron charging phenomena, the resistance of each tunnel barrier isolating the studied object should be much larger than the quantum resistance ($R_q = h/e^2$), which ensures that the wavefunction of an excess electron on the studied object is well localized there. Previously, it was believed that single-electron charging states should not be observed in a continuous graphene sheet and, definitely, they have not been reported yet. Therefore, the observation of single-electron charging effects in the GQDs of the continuous graphene sheet, as reported in this work, provides direct evidence that the Dirac fermions can be trapped in partial regions of the graphene monolayer.

In our experiment, graphene monolayer was grown on molybdenum foils using atmospheric pressure chemical vapor deposition method [17,18]. Firstly, transition-metal carbide (TMC), such as $Mo_2C$, was formed from parent metal foils. Then, thus-formed early $Mo_2C$ served as catalysts for graphene growth (details are described in the Supplementary Information [19], and see Fig. S1 and S2 for other characterizations of the sample, and see Fig. S3 for the growth process). The STM investigation of the as-grown samples revealed the presence of reconstructed $Mo_2C$ surface underneath the continuous graphene monolayer. These reconstructions of the TMC surface consist of nanoscale islands and quantum-dot-like vacancy islands (see Fig.1a and Fig. S3 [19]). The graphene monolayer is suspended over the vacancy islands of the surface and the van der Waals forces from the STM tip can induce substantial mechanical deformation in the nanoscale suspended graphene membranes [20,21]. Our experiment indicates that the suspended graphene nanomembrances over the vacancy islands show a reversible mechanical deformation in response to the change of tip-graphene distances, as shown in Fig. 1b. There may be strong coupling between Mo atoms around the boundaries of the vacancy islands and the carbon atoms of graphene, which leads to formation of boundaries with nanoscale width [17,18,22,23]. Additionally, the charge transference between graphene and the TMC surface is expected to be affected by spatial variation of distance between graphene and TMC surface around the boundaries of the vacancy islands [24]. These two effects can result in electron confinement in partial regions of a continuous graphene sheet. We will show subsequently that the suspended graphene nanomembrances over the vacancy islands behave as isolated GQDs (from now on, we use GQDs to refer to the suspended graphene regions over the vacancy islands in this work).

Our scanning tunneling spectroscopy (STS) measurements, as shown in Fig. 1c, indicates that the suspended graphene region exhibits quite different electronic properties comparing to the surrounding graphene sheet. Inside the suspended graphene region we observe a series of almost equally spaced resonances in the tunneling conductance. Outside such a suspended graphene region we observe a V-shaped spectrum, as expected to be observed for graphene monolayer on metallic surface. Similar phenomena have also been observed around other suspended graphene regions in our experiment. The spatial variation of the tunneling spectra, as shown in Fig. 1c and Fig. S10 [19], precludes any possible artificial effects of the STM tips as the origin of these features. We attribute the peaks in the tunneling spectra recorded inside the suspended graphene regions to Coulomb oscillations, which are expected to be observed in the tunneling conductance of the isolated GQDs [10]. Such a result indicates that each suspended graphene region over the vacancy island is electronically isolated from the surrounding graphene sheet by an insulating barrier and behaves as a GQD. When the STM tip is positioned above a GQD, an asymmetric double-barrier tunnel junction (DBTJ), as schematically shown in Fig. 2a, is formed. One of the tunnel barrier, described by a capacitor $C_T$ in parallel with an ohmic resistor $R_T$, is generated between the STM tip and the GQD; the other tunnel barrier, described by a capacitor $C_B$ in parallel with an ohmic resistor $R_B$, is generated between the GQD and the surrounding graphene sheet. Therefore, the tunneling spectra of the GQDs can be tuned by varying the tip-sample distances and such a behavior should be described well by the orthodox Coulomb blockade theory [6].

To verify the above assumption, we measured the tunneling spectra of the GQD by varying the tip-sample distances, which can be realized by tuning the bias voltages or changing the tunneling currents during the STM measurement (see Fig. S5 [19]). The variation of the tip-sample distances changes the tunnel resistance $R_T$ dramatically and, therefore, affects the spectra of the GQD according to the effective circuit of the DBTJ. Figure 2b shows a representative result obtained in the GQD (see Fig. S5 for more experimental data [19]). For large tip-sample distance (large $R_T$), the spectra exhibit the typical signature of Coulomb blockade (CB), i.e., a zero-conductance gap around the Fermi energy. For small tip-sample distance (small $R_T$), the spectra present quasi-periodic tunneling peaks, known as the signature of Coulomb staircase (CS). Obviously, evolution of the spectra between the CB regime and the CS regime is observed, as shown in Fig. 2b, by a controlled change of the resistance $R_T$, i.e., the distance between the STM tip and the GQD. Such a result can be reproduced well by the optimized simulation based on the orthodox Coulomb blockade theory [6], as shown in Fig. 2c. In the simulation, $R_B$ is assumed to be a constant and we only change the value of $R_T$ according to the experimental condition (some factors determining the shape and intensity of conductance peaks are taken into account, see Supplementary Information for details [19]). The consistency between the experimental data and the theoretical results demonstrates explicitly that the suspended graphene region over the vacancy island behaves as an isolated GQD.

Here we should point out that the tunneling spectra of the nanoscale GQDs, as shown in Fig. 2b, also exhibit other two important features beyond the description of the orthodox Coulomb blockade theory. The first one is the relatively wide distribution of the energy spacing between the nearest-neighbor tunneling peaks of the spectra. Figure 2d shows representative histograms of the nearest-neighbor level spacing of three GQDs observed in our experiment. The random tunneling peak spacing may arise from the roughness and irregular geometry of GQDs, which can be well described by the theory of chaotic neutrino billiards [10]. Moreover, the quantum confinement effect can also influence the tunneling peak spacing for nanoscale GQDs, and may alter the fitting parameters of the DBTJ model. However, the quantum confinement effect cannot change the qualitative features of Coulomb oscillation in the GQDs. Considering that the quantitative description of quantum confinement effect relays on detailed information of GQD boundaries, which is beyond current experimental condition, we ignore the effect of quantum confinement in the theoretical simulations. The second feature beyond the description of the orthodox theory is the deviation between the histograms of the nearest-neighbor level spacing for electrons and holes, as shown in Fig. 2d. There is a slight difference between the maximum of the histograms for electrons and holes. Figure 2e shows a representative result about energy positions of the peaks as a function of the integer number obtained in a tunneling spectrum of a GQD (similar result has been observed in all the spectra of the GQDs in our experiment). According to the slopes of the data, we can conclude that there is a notable difference between the average energy spacing of the nearest-neighbor tunneling peaks for electrons and holes. Such a behavior, which has never been reported in previous single-electron charging effects, is attributed to electron-hole asymmetry in graphene monolayer. The existence of electron-hole asymmetry in graphene has been demonstrated through Landau level spectroscopy [9] and transport measurement [30,31] previously. To better describe the experimental data using the orthodox theory, we assume different values of the capacitor $C_B$ for electrons and holes, i.e., we have $C_B^e \neq C_B^h$ (consequently, we have different residual charges for electrons and holes, i.e., $Q^e \neq Q^h$), to account for the electron-hole asymmetry in graphene. A moderate difference between the capacitor $C_B$ and the residual charges for electrons and holes well describes the observed electron-hole asymmetry in the tunneling spectra of the GQDs, as shown in Fig. 2f.

The electronic properties of the GQDs are further studied by operating energy-fixed STS mapping. Figure 3a and 3b shows two representative STS maps at different energies, which exhibit striking concentric rings of differential conductance peaks inside the GQDs. Each ring in the STS maps corresponds to a single Coulomb oscillation of the GQD [14,15]. When the STM tip is moved above the GQD, it induces spatial variation of band-bending in the GQD [32,33], as schematically shown in Fig. 3c. This tip-induced gating leads to maxima in the measured STS maps at certain positions, where the Fermi level cross one of the tunneling peaks in the spectra. The gate-dependent band-bending mechanism explains the nearly concentric rings seen in each GQD in the conductance maps. Such a result further demonstrates that the suspended graphene regions over the vacancy islands (the GQDs) are electronically isolated from the surrounding graphene sheet.

It is now natural to ask how the electrons are confined in the GQDs in the continuous graphene sheet. The spatial variation of distance between graphene and TMC surface around the boundaries of the vacancy islands can generate nanoscale circular pn junction[21] in the graphene sheet (see Fig. S8 [19]). To explore the effect of nanoscale circular pn junction on the electronic properties of the GQDs in the graphene sheet, we studied a similar structure, i.e., a GQD in a continuous graphene sheet, grown on a Cu foil for comparison. The electrons are temporarily trapped inside the circular pn junction to form the quasi-bound states [4-6], as revealed by a series of resonance peaks at negative energies in the tunneling spectra (see Fig. S14 in Supplementary Information for experimental data and see Fig. S15 for our simulation results [19]). Both experiment and theory demonstrate that the lowest resonance of the quasi-bound states exhibits maxima near the centre of the GQD, whereas higher resonances display stronger intensity close to the boundary of the GQD (see Fig. S14 and S15 [19]), as reported previously in Ref. 6. Obviously, the main features of the single-electron transport in the GQD on TMC surface, as shown in Fig. 2 and Fig. 3, are different from that of the quasi-bound states inside the circular pn junction on Cu surface, as shown in Fig. S14. We attributed the origin of the difference to the different graphene-substrate coupling: graphene is strongly chemisorbed on TMC, whereas the binding to Cu is much weaker. The existence of nanoscale boundaries where the carbon atoms of graphene are strongly coupled with Mo atoms around the boundaries of the vacancy islands may play a critical role in the electron confinement in the GQDs.

To further explore the origin of the tunnel barrier between the GQDs and the surrounding continuous graphene sheet, we measured atomic-resolved STS spectra around the boundary of the GQD by fixing the tip-sample distance (the resistance $R_T$), as shown in Fig. 4a and 4b. When approaching the boundary from the inside of the GQD, we observed evolution of the spectra from the CS regime to the CB regime. This indicates that the $R_B$ decreases dramatically from position *i* to position *iii* (see Fig. 4c). The most striking result is that the spectrum recorded at position *iv* becomes V-shaped, as expected to be observed in the continuous graphene sheet. Such a result indicates that the effective boundary of the GQD spreads over position *i* to position *iii*，with typical width of several nanometers. In graphene, the $\pi$ electrons are responsible for the electronic properties at low energies. Owing to the strong graphene-Mo interaction, the $\pi$ orbital of the graphene is hybridized with the *d* orbital of the Mo atoms within the boundary regions and these $\pi$ electrons around the boundary become strongly localized (see illustration in Fig. S3). Therefore, the electrons inside the GQD are strongly confined. The confinement of electrons due to the tunnel barrier between GQD and continuous graphene sheet is further demonstrated through imaging the intervalley scattering around the boundary of the GQD, as shown in Fig. 4d. The existence of valley mixing is seen as the emergence of both the $\sqrt{3}\times\sqrt{3}R30°$ interference pattern of carbocyclic rings in the STS maps and the inner six spots in the fast Fourier transform image [34,35]. Although further analysis is necessary, our results indicate that the atomically sharp boundaries induced by strong graphene-substrate interaction are very important in the electron confinement in the GQDs.

In summary, bound states are realized for the first time in nanoscale GQDs in the continuous graphene sheet. Our result indicates that strong graphene-substrate coupling plays a vital role in confining electrons in the nanoscale GQDs. The method reported in this work may pave a new road to electronically isolate graphene nanostructures in a continuous graphene monolayer, which is very important in advanced graphene nanoelectronics.

**Acknowledgements**

This work was supported by the National Basic Research Program of China (Grants Nos. 2014CB920903, 2013CBA01603, 2014CB920901), the National Natural Science Foundation of China (Grant Nos. 11674029, 11422430, 11374035, 11374219, 11504008), the program for New Century Excellent Talents in University of the Ministry of Education of China (Grant No. NCET-13-0054). L.H. also acknowledges support from the National Program for Support of Top-notch Young Professionals.

# Figures

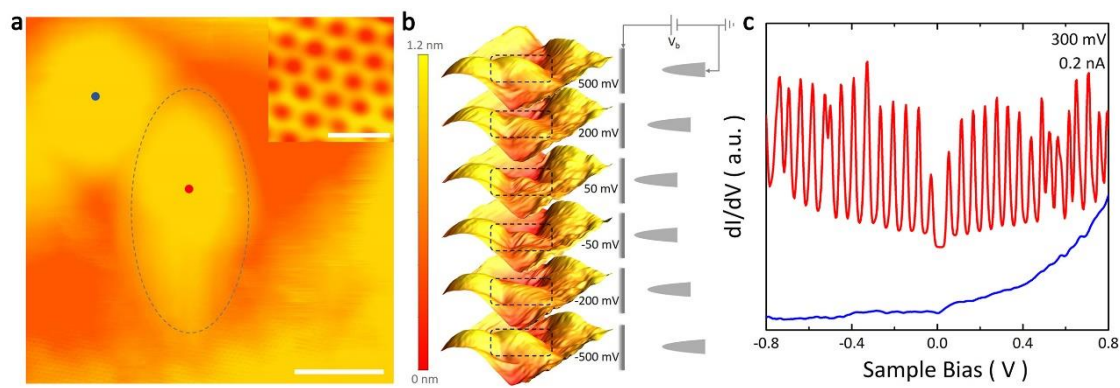

**Figure 1 |** STM and STS measurements around a GQD. **a,** A STM topographic image ($V_b$ = 500 mV, $I_S$ = 0.20 nA) showing a suspended graphene nanostructure over a vacancy island of the substrate, as marked with the ellipse-shaped dashed curve. The scale bar is 5 nm. The inset shows atomic structure of graphene obtained in the studied region. The scale bar in the inset is 0.5 nm. The suspended graphene nanostructure behaves as an isolated GQD. **b,** STM topographic images of the suspended graphene nanostructure over the vacancy island for different bias voltages (i.e., for different tip-sample distances). **c,** the typical dI/dV spectra recorded inside and outside of the suspended graphene nanostructure (the GQD). The curves are offset in Y-axis for clarity. a.u., arbitrary units.

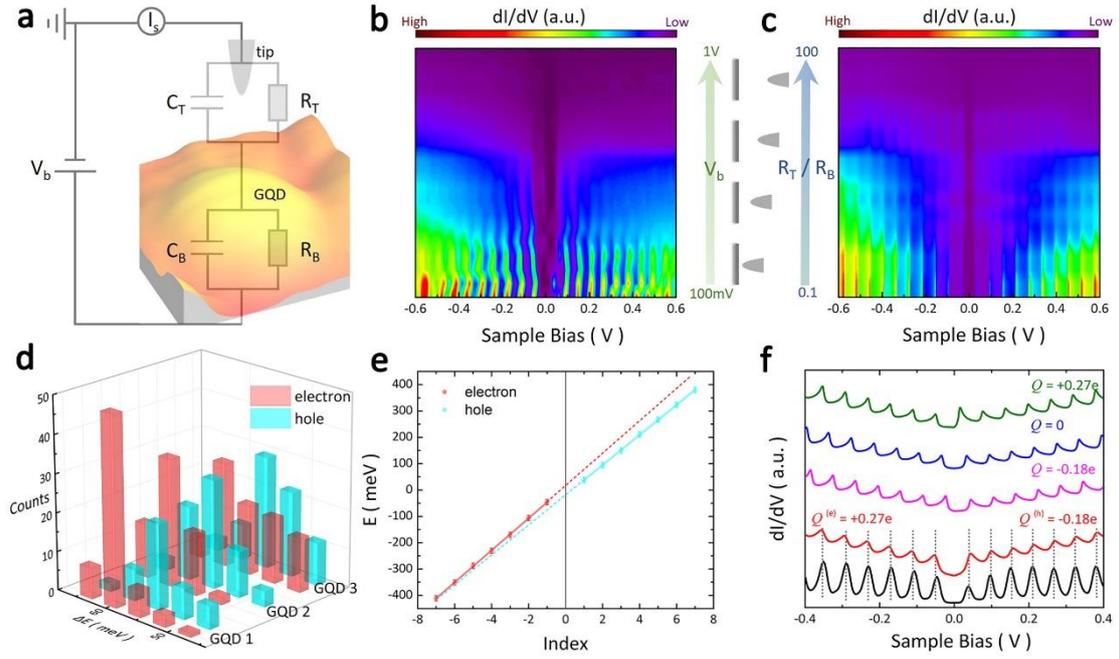

**Figure 2 |** Spectra of single-electron effects in the GQDs. **a,** A DBTJ circuit where each junction is represented by a set of capacitors and resistors. The semi-transparent background shows the schematic diagram of the single-electron charging system, in which the GQD is electronically isolated from the surrounding continuous graphene sheet. **b** shows representative STS spectra of a GQD by varying the bias voltages, i.e., the tip-sample distance. **c** shows simulated tunneling spectra of a GQD by changing the ratio of $R_T/R_B$. The parameters used in the calculation are $C_B = 2.67$ aF, $C_T = 0.1$ aF, $R_B = 1$ MΩ, and $Q = 0$. **d,** 3D histogram of the nearest-neighbor level spacing of the conductance peaks in the STS spectra acquired on different GQDs. **e,** Energy levels of conductance peaks, extracted from the spectra in the d$I$/d$V$ map shown in panel **b** at $V_b = 110$ mV, as a function of the peak index (electron, negative integer numbers; hole, positive integer numbers). The linear fitting of the data indicates a slight deviation of the average energy spacing of the nearest-neighbor peaks for electron and hole. **f,** A typical STS spectrum of the GQD recorded at $V_b = 110$ mV (black line). The red curve

is the simulated result by taking into account the electron-hole asymmetry. In the calculation, we used $C_B^e = 2.67$ aF, $C_B^h = 2.85$ aF, $Q^e = 0.27e$, and $Q^h = -0.18e$. The green, blue and pink curves are also simulation results but without considering the electron-hole asymmetry. The remaining parameters in the calculation are $R_B = 1 M\Omega$, $R_T = 0.2 M\Omega$, and $C_T = 0.1$ aF.

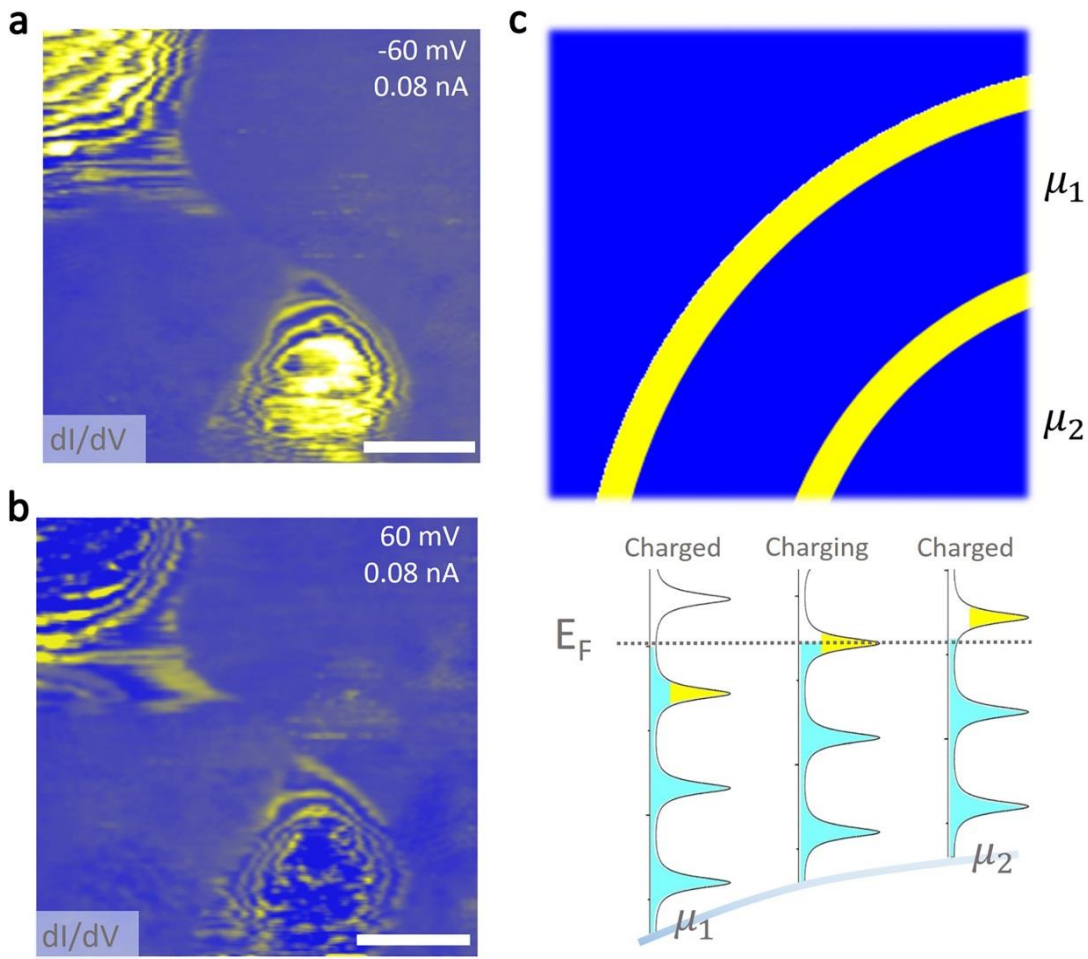

**Figure 3 |** Imaging the single-electron states in the GQDs. **a and b,** dI/dV maps recorded around the GQDs with the fixed sample bias of -60 mV (a) and 60 mV (b), respectively. The scale bar is 5 nm. There are two sets of concentric rings of conductance peaks, centered at two different GQDs. Each ring in the STS maps

corresponds to a single Coulomb oscillation on the GQD. **c,** A schematic diagram showing effects of tip-induced band bending on the formation of the concentric rings inside the GQDs in the conductance maps.

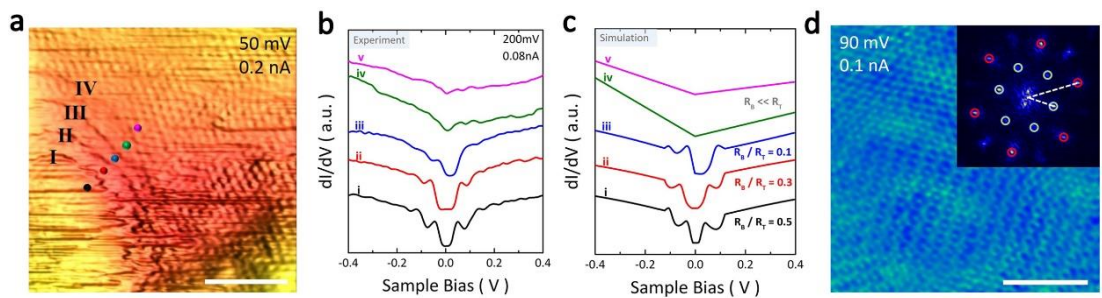

**Figure 4 |** STS spectra and valley mixing around the atomically sharp boundary of the GQD. **a,** Zoom-in atomic-resolution topographic images obtained at the top right corner of the GQD. The scale bar is 2 nm. **b,** STS spectra recorded at different positions in panel **a** (indicated by colored dots) by fixing the tip-sample distance. **c,** simulated tunneling spectra of a GQD by changing the ratio of $R_B/R_T$. In the simulation, $R_T$ is fixed according to the experiment. The dramatically changes of the spectra recorded at positions *i* and *iv* indicate that the effective width of the boundary of the GQD is about 2 nm. **d,** A representative STS map recorded around the boundary of the GQD. A clear interference pattern arise from the intervalley scattering is observed. The inset shows fast Fourier transform of the STS map. The outer six spots are the reciprocal lattice of graphene and the inner six spots arise from valley mixing induced by the boundary of the GQD.

18